\documentclass[aps,preprint,superscriptaddress]{revtex4-1}
\usepackage{amsmath,bm,amssymb,amsfonts}
\usepackage{dcolumn}
\usepackage{graphicx}
\usepackage{latexsym}
\usepackage{epsfig}
\usepackage{multirow}

\usepackage{color}

\begin{document}

\title{Topological band gap in intercalated epitaxial graphene}

\author{Minsung \surname{Kim}}
\affiliation{Ames Laboratory -- U.S. Department of Energy, Iowa State University, Ames, Iowa 50011, USA}
\author{Cai-Zhuang \surname{Wang}}
\affiliation{Ames Laboratory -- U.S. Department of Energy, Iowa State University, Ames, Iowa 50011, USA}
\affiliation{Department of Physics and Astronomy, Iowa State University, Ames, Iowa 50011, USA}
\author{Michael C. \surname{Tringides}}
\affiliation{Ames Laboratory -- U.S. Department of Energy, Iowa State University, Ames, Iowa 50011, USA}
\affiliation{Department of Physics and Astronomy, Iowa State University, Ames, Iowa 50011, USA}
\author{Myron \surname{Hupalo}}
\affiliation{Ames Laboratory -- U.S. Department of Energy, Iowa State University, Ames, Iowa 50011, USA}
\affiliation{Department of Physics and Astronomy, Iowa State University, Ames, Iowa 50011, USA}
%\author{Patricia A. \surname{Thiel}}
%\affiliation{Ames Laboratory -- U.S. Department of Energy, Iowa State University, Ames, Iowa 50011, USA}
%\affiliation{Department of Chemistry, Iowa State University, Ames, Iowa 50011, USA}
%\affiliation{Department of Materials Science and Engineering, Iowa State University, Ames, Iowa 50011, USA}
\author{Kai-Ming \surname{Ho}}
\affiliation{Ames Laboratory -- U.S. Department of Energy, Iowa State University, Ames, Iowa 50011, USA}
\affiliation{Department of Physics and Astronomy, Iowa State University, Ames, Iowa 50011, USA}
%\author{aaa \surname{aaa}}
% \email{To whom correspondence may be addressed. E-mail: jihm@snu.ac.kr, sbchung@snu.ac.kr}
%\affiliation{Department of Physics and Astronomy, Seoul National University, Seoul 151-747, Korea}

\date{\today}

\begin{abstract}
Functional manipulation of graphene is an important topic 
in view of both fundamental researches and practical applications.
In this study, we show that intercalation of 5$d$ transition metals in epitaxial graphene on SiC 
is a promising approach to realize topologically nontrivial phases with a finite band gap in graphene.
Using first-principles calculations based on density functional theory, 
we show that the Re- and Ta-intercalated graphene become 
two-dimensional topological insulators which exhibit linear Dirac cones and quadratic bands
with topological band gaps, respectively.
The appearance of the topological states is attributed to 
the strong spin-orbit coupling strength of the intercalants. 
%which gives rises to topologically nontrivial finite gaps.
We find that topological edge states exist within the finite bulk band gap 
in accordance with the bulk-boundary correspondence.
We also discuss the spin splitting of the band structure 
due to the inversion symmetry breaking and the spin-orbit coupling.
Our results demonstrate that intercalation of graphene is an effective and viable method 
to manipulate the band gap and the topological character of graphene.
Such intercalated graphene systems are potentially useful 
for spintronics and quantum computing applications. 
\end{abstract}

%\pacs{ }

\date{\today}

\maketitle

%For the two-dimensional electron gas (2DEG) at the surfaces and interfaces of the perovskite transition metal (TM) oxide
%~\cite{Ohtomo2004,Takagi2010,Mannhart2010,Santander2011,Meevasana2011},
%the recent years have seen much debate on the magnitude of the spin-orbit interaction
%~\cite{BenShalom2010,Caviglia2010,Santander2014,Santander2012,King2012,Reyren2012,MKim2012}.
%One feature 

%\emph{Introduction}---
Low-dimensional material has been a primary theoretical and experimental interest 
in condensed matter physics, chemistry, and materials science 
since the first synthesis of graphene~\cite{novoselov2004,novoselov2005,zhang2005}. 
One of the most important aspects of this two-dimensional (2D) atomically-thin carbon layer is that 
it allows functional manipulation of electronic properties 
via physical or chemical interactions with the substances adjacent to it, 
e.g., using molecular adsorption, heterostructure, or intercalation.
In particular, intercalation was shown to be a useful method 
to affect the quasi-particle excitation, superconductivity, adsorption energy of adatoms, 
etc.~\cite{starke2012,gierz2010,emtsev2011,ludbrook2015,ichinokura2016,schumacher2013}.
It was demonstrated that
the intercalation controls the coupling between a substrate 
and a graphene layer, and can further change the band character of Dirac cones in  
graphene~\cite{gierz2010,emtsev2011,starke2012,riedl2009,mskim2017,li2012}.

Among many intriguing electronic properties of graphene, 
the quantum spin Hall (QSH) phase (i.e., the 2D topological insulator phase) 
attracted noticeable interest~\cite{hasan2010,qi2011}.
Theoretically, the carbon honeycomb lattice is expected to show 
the QSH effect characterized by the nontrivial $\mathbb{Z}_2$ 
topological number~\cite{kane2005qu,kane2005z2}.
However, the essential ingredient of topological phases is spin-orbit coupling (SOC)
which is small in graphene~\cite{yao2007}.
The SOC-induced band gap of pristine graphene is estimated to be $10^{-3}$ meV
which is negligible in practice.
Since the gapless band structure sets fundamental limitation in the application of graphene
in comparison with other conventional semiconductors with finite gaps,
the enhancement of the SOC strength is important for the utility of graphene as well as
the realization of the novel topological phase.  
%It is plausible that the SOC strength can be enhanced by intercalation
%of another layer of heavy elements.

In this study, we consider the intercalation of 5$d$ transition metal atoms that has 
significant SOC strength to enhance the topological band gap of graphene.
Given that the intercalation can give rise to qualitatively different Dirac cones
with significant contributions from intercalant orbital states~\cite{li2012},
it is plausible to expect that the intercalation of such heavy elements would
enhance the SOC effects in the Dirac cones. 
After a systematic calculation of the electronic band structures 
of the epitaxial graphene on SiC substrate with 5$d$ transition metal
(from Hf to Ir in the periodic table) intercalation, 
we find that 2D topological insulator phases
are indeed achieved for suitable choices of the intercalants.
%In this study, we propose the intercalated epitaxial graphene on SiC 
%as a promising candidate material for the realization of topological phases
%in graphene system.
%Specifically, we consider the intercalation of $5d$ transition metals Re and Ta 
%which have considerable sizes of the SOC strength.
%Using first-principles electronic band structure calculations, 
Specifically, we show that the Re-intercalated epitaxial graphene 
has topological Dirac cones at the Fermi energy.
We find that the SOC induces a finite topological band gap with 
nontrivial $\mathbb{Z}_2$ band topology of the occupied bands.
Correspondingly, the spin-polarized topological edge states appear 
due to the nontrivial 2D bulk band topology as dictated by bulk-boundary correspondence.
In the case of the Ta intercalation, the electronic band structure near the Fermi level
shows quadratic band dispersion (as opposed to linear one in the Re intercalation) 
with a topological band gap.
We also show that the spin splitting of the band structures near the band edges 
is originated from the broken inversion symmetry and the SOC.

The electronic band structures were calculated using a first-principles method based on
density functional theory (DFT) as implemented in VASP package~\cite{kresse1993,kresse1996}.
We employed the projector augmented-wave (PAW) method~\cite{blochl1994,kresse1999} 
and used a plane-wave basis set with 400 eV energy cutoff.
$\Gamma$-centered $12\times12\times1$ $k$-point meshes were exploited.
For the atomic structure, we considered $\sqrt{3}\times\sqrt{3}$ cell 
of the Si-terminated SiC substrate (consisting of 4 SiC layers)
which may accommodate $2\times2$ graphene layer on top of it.
An experimental lattice constant of SiC (type 6H) was used~\cite{gomes1967}, and
the internal atomic coordinates were relaxed until the force acting on each atom
became less than 0.02 eV/\AA ~while the 2 lowest SiC layers were fixed.
The Re and Ta metal atoms were intercalated between the outermost Si layer
and the graphene layer (i.e., decoupled buffer layer).
A sufficiently thick vacuum region ($\gtrapprox$ 20 \AA) was employed 
to prevent the unwanted interaction between periodic images.
Wannier functions were obtained using Wannier90~\cite{mostofi2014} and
edge state dispersions were calculated using WannierTools~\cite{wu2018}.

\begin{figure}[]
\includegraphics[width=0.48\textwidth]{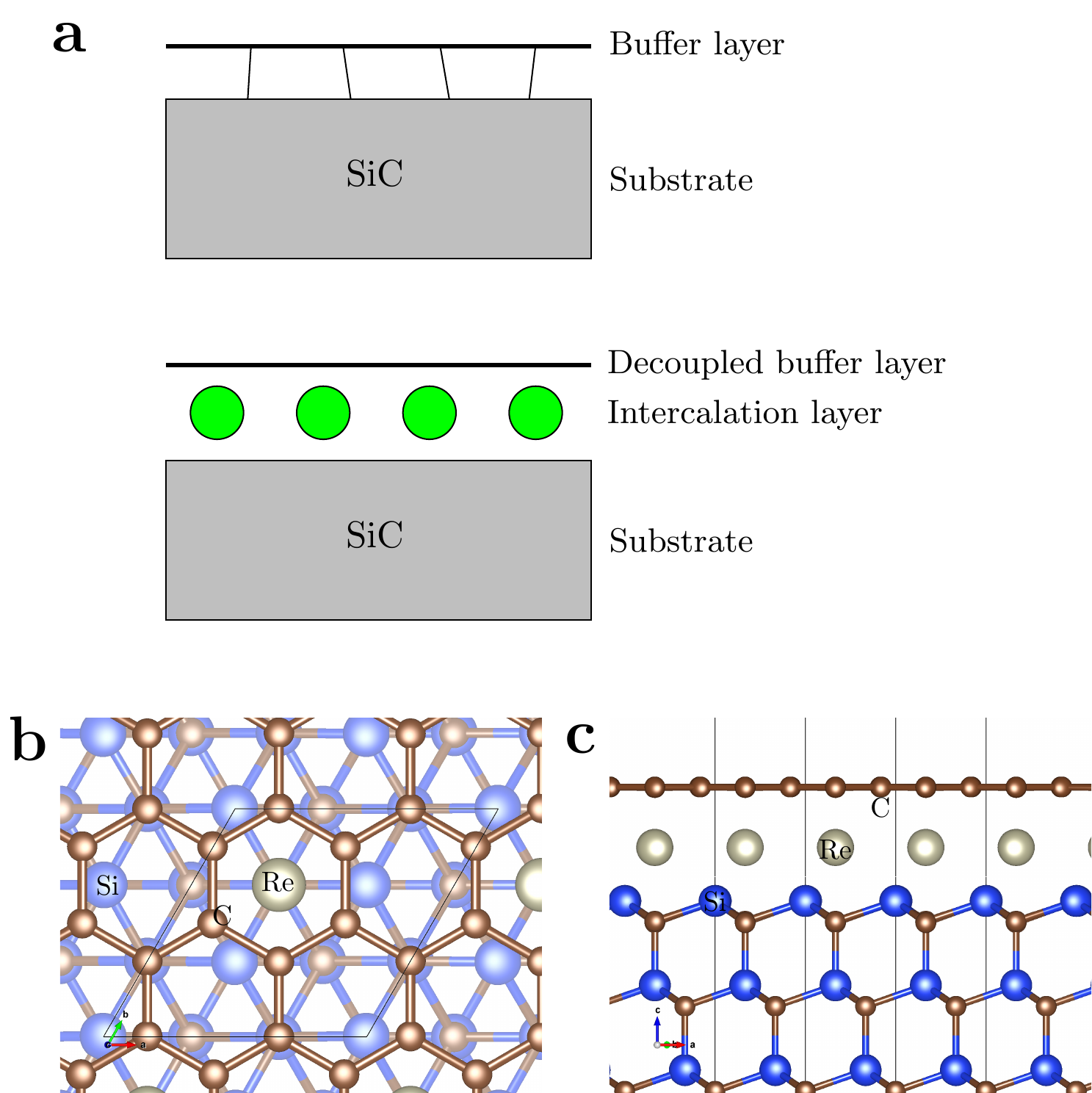}
\caption{\label{fig:re_intercal_atmstr} Atomic structure of Re-intercalated epitaxial graphene on SiC. (a) Schematic illustration. (b) Top and (c) side view of the atomic structure.
}
\end{figure}

The intercalation induces decoupling of the graphene layer from the SiC substrate.
In the epitaxial graphene on SiC substrate, the first carbon layer (called buffer layer or 0th layer) 
makes strong chemical bonds with the outermost Si atoms.
Due to the strong hybridization, 
this buffer layer does not possess Dirac cones 
unlike pristine graphene~\cite{gierz2010,emtsev2011}.
The intercalation can decouple the buffer layer from the substrate reviving 
the Dirac cones of the graphene (i.e., decoupled buffer layer)
~\cite{gierz2010,watch2013,watch2012,emtsev2011,starke2012,riedl2009,mskim2017}.
For instance, the intercalation of Au was experimentally shown to control
the coupling between the buffer layer and the substrate as demonstrated by
the appearance of Dirac cones upon intercalation 
using angle-resolved photoemission spectroscopy (ARPES)~\cite{gierz2010}.
In our present study the intercalation of the Re or Ta layer (Fig.~\ref{fig:re_intercal_atmstr}) 
is also expected to decouple the buffer layer from the SiC substrate and 
change the effective SOC strength in the graphene layer.

\begin{figure}[]
\includegraphics[width=0.48\textwidth]{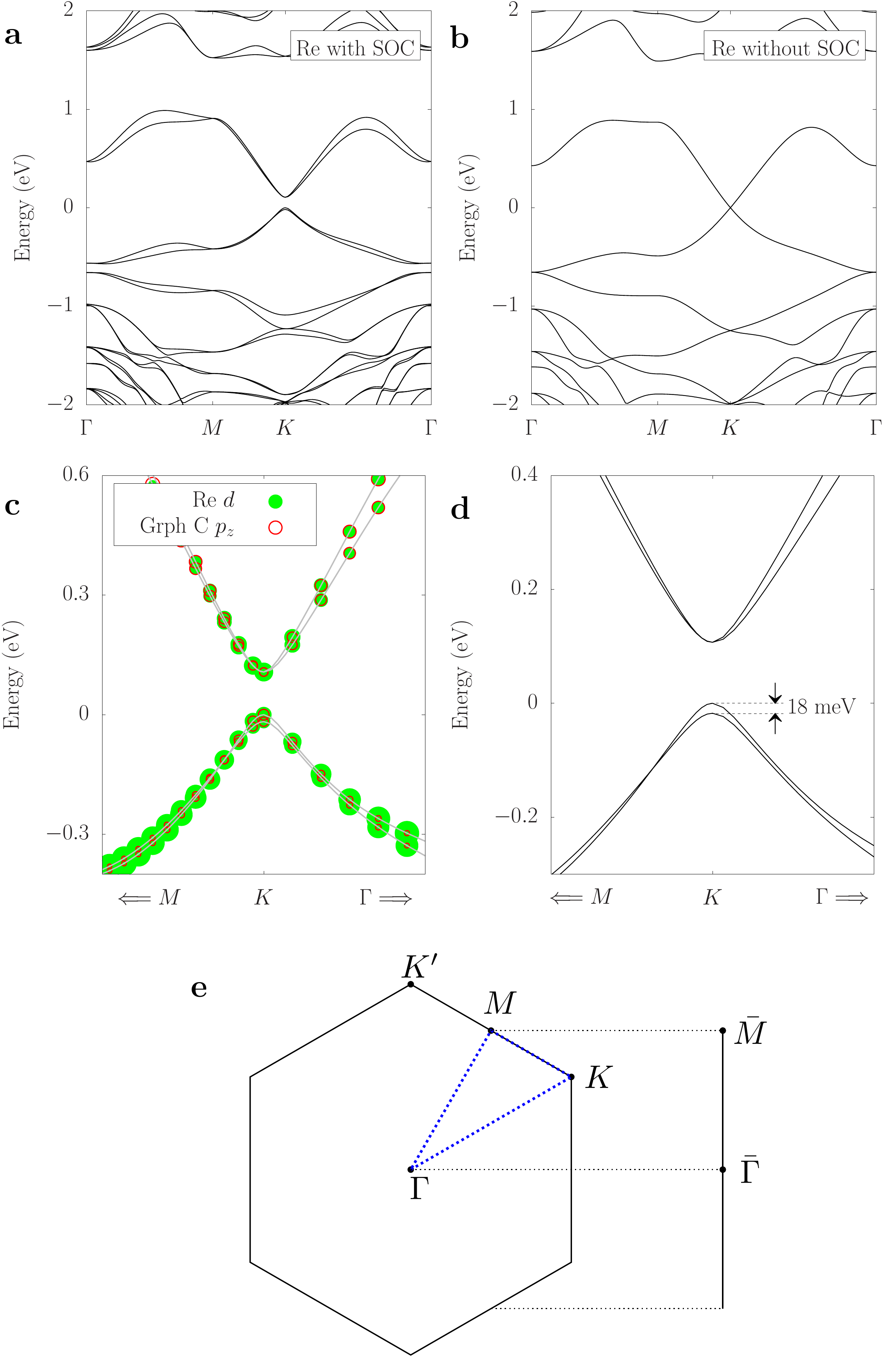}
\caption{\label{fig:re_intercal_bndstr} Electronic band structure of Re-intercalated epitaxial graphene on SiC. 
The band structure (a) with and (b) without SOC. 
(c) Contributions from Re $d$ and graphene C $p_z$ states are denoted by green filled and red open circles, respectively.
(d) The band structure near $K$ with SOC.
(e) The corresponding BZ.
}
\end{figure}

We find that the electronic band structure of the Re-intercalated epitaxial graphene has 
Dirac cones with a sizable SOC-induced band gap.
Figure~\ref{fig:re_intercal_bndstr} shows the electronic band structures
obtained from our DFT calculations.
Without the SOC, the band structure has a Dirac cone at the $K$ point of the Brillouin zone (BZ) 
as in pristine graphene.
However, upon including the SOC, 
there appears a finite band gap at the Dirac point ($\approxeq$ 107 meV) 
in contrast to pristine graphene.
To understand the origin of the SOC-induced gap, we calculate the orbital contributions
of the Dirac cone states (Fig.~\ref{fig:re_intercal_bndstr}c).
In pristine graphene, the Dirac cone at the $K$ point consists of
C $p_z$ orbital.
In contrast, in the epitaxial graphene with the Re intercalation, 
we find that there are significant contributions from Re $d$ states 
as well as the graphene C $p_z$.
Thus, we have the sizable SOC-induced gap opening in the Re-intercalated epitaxial graphene
whereas the effect of SOC on the band gap is negligible in pristine graphene.

To examine the topological feature of the intercalated epitaxial graphene,
we calculate the $\mathbb{Z}_2$ topological invariant 
that characterizes the band topology of a 2D gapped system~\cite{fu2006}.
The topological number is defined as
\begin{eqnarray}
\nu = \frac{1}{2\pi} \left\{ \oint_{\partial B_{1/2}} A -\int_{B_{1/2}} F \right\} ~\mathrm{mod}~2,
\end{eqnarray}
where $A$, $F$, and $B_{1/2}$ represent the Berry connection, the Berry curvature, 
and the half of the BZ, respectively.
In practice, this integral formula is computed on a discretized $k$-mesh in DFT calculations and
the corresponding topological number becomes a sum of an integer field $n(\vec{k})$ as 
$\nu = \sum_{\vec{k}_i \in B_{1/2}} n(\vec{k}_i)~\mathrm{mod}~2$~\cite{fukui2007,xiao2010}.
Whereas the topological Chern number can have any integer value
(which physically corresponds to the number of edge states in the quantum Hall system),
the topological number for the QSH classification is defined only up to mod 2
due to the presence of the time-reversal symmetry~\cite{fu2006}.
We find that the $\mathbb{Z}_2$ invariant of the occupied bands 
in the Re-intercalated epitaxial graphene is nontrivial, i.e., $\nu = 1$.
Thus, the Re-intercalated graphene is 
a 2D topological insulator (or a QSH system).

Another notable effect of the SOC is spin degeneracy lift in the electronic band structure.
In pristine graphene, the band structure is spin-degenerate at each $k$-point of the BZ
due to the simultaneous presence of the spatial inversion symmetry and
the time reversal symmetry as dictated by the relation
$E_{\vec{k}\uparrow}=E_{-\vec{k}\uparrow}=E_{\vec{k}\downarrow}$,
where the first and the second equality is given by the inversion and the time-reversal
symmetry, respectively.
If the inversion symmetry is broken, 
the spin degeneracy will be lifted by the SOC in general.
In the intercalated epitaxial graphene, the inversion symmetry is broken
due to the presence of SiC substrate and the intercalation layer. 
Thus we have the spin splitting upon including the SOC.
%in the presence of the SOC. 
The size of the spin splitting in the valence band maximum (VBM) at $K$ is
calculated to be $\approxeq$ 18 meV (Fig.~\ref{fig:re_intercal_bndstr}d).
We note that a monolayer of the transition metal dichalcogenide 
MoS$_2$ shares important common features with the intercalated graphene
in that both systems have honeycomb-derived lattices, broken inversion symmetry, 
and sizable SOC strengths.
We note that the monolayer MoS$_2$ has a direct gap at $K$ 
with larger VBM spin splitting of $\approxeq$ 0.148 meV~\cite{zhu2011}.

\begin{figure}[]
\includegraphics[width=0.48\textwidth]{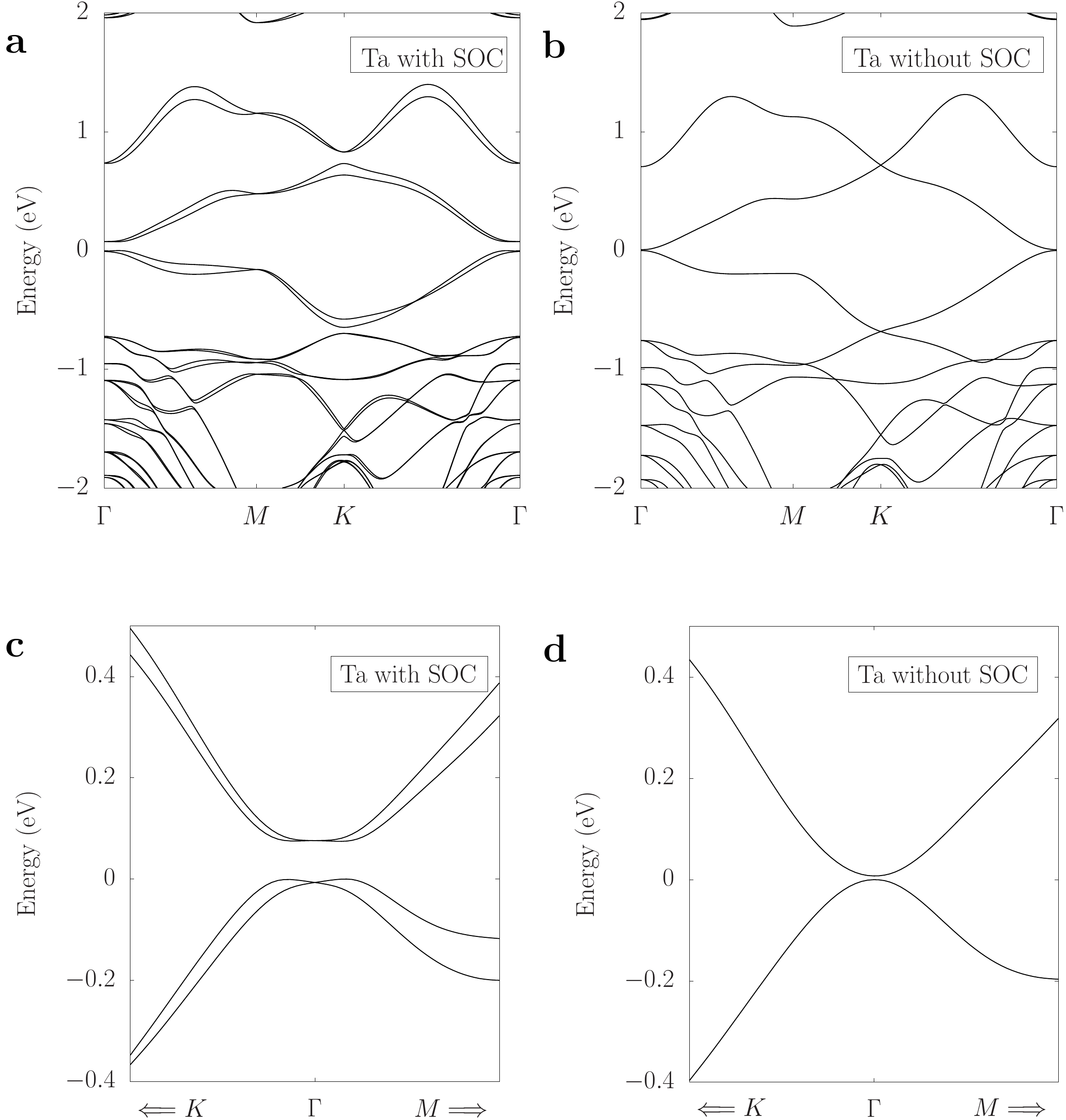}
\caption{\label{fig:ta_intercal_bndstr} Electronic band structure of Ta-intercalated epitaxial graphene. 
The band structure (a) with and (b) without SOC. 
The band structure near $\Gamma$ (c) with and (d) without SOC.
}
\end{figure}

The Ta-intercalated epitaxial graphene also shows a 2D topological insulator phase.
The main difference in comparison with the Re intercalation is that
the electronic band dispersion is quadratic (as opposed to linear) 
in the absence of the SOC (Fig.~\ref{fig:ta_intercal_bndstr}).
Since Ta atom has 2 fewer electrons than Re, 
the Fermi energy shifts downward by two bands (including spin) compared with the Re intercalation,
and the VBM and the conduction band minimum (CBM) lie
near the $\Gamma$ point of the BZ.
The $\mathbb{Z}_2$ invariant of the occupied bands is calculated to be 1, namely,
the Ta-intercalated epitaxial graphene is also a 2D topological insulator
(with a gap of $\approxeq$ 74 meV).
Unlike the $K$ point, the $\Gamma$ point is a time-reversal invariant momentum
satisfying $-\vec{k}=\vec{k}$, 
hence the spin degeneracy at $\Gamma$ is protected by the time-reversal symmetry.
Thus, the band dispersion near $\Gamma$ shows the Rashba-type spin splitting
(Fig.~\ref{fig:ta_intercal_bndstr}c)~\cite{bychkov1984,lashell1996}. 
The strength of the Rashba-type spin splitting can be represented by
the Rashba parameter $\alpha_R$ which is defined 
through the Bychkov-Rashba Hamiltonian
\begin{eqnarray}
\mathcal{H}_{R} = \alpha_R (\hat{\gamma} \times \vec{k}) \cdot \vec{\sigma},
\end{eqnarray}
where $\hat{\gamma}$ and $\vec{\sigma}$ denote the direction of 
the inversion symmetry breaking field (that is perpendicular to the graphene plane in our case) 
and Pauli matrices, respectively.
According to our DFT band structure, 
the magnitude of the Rashba parameter $\alpha_R$ in the valence band
is estimated to be 0.21 eV\AA.

\begin{figure}[]
\includegraphics[width=0.46\textwidth]{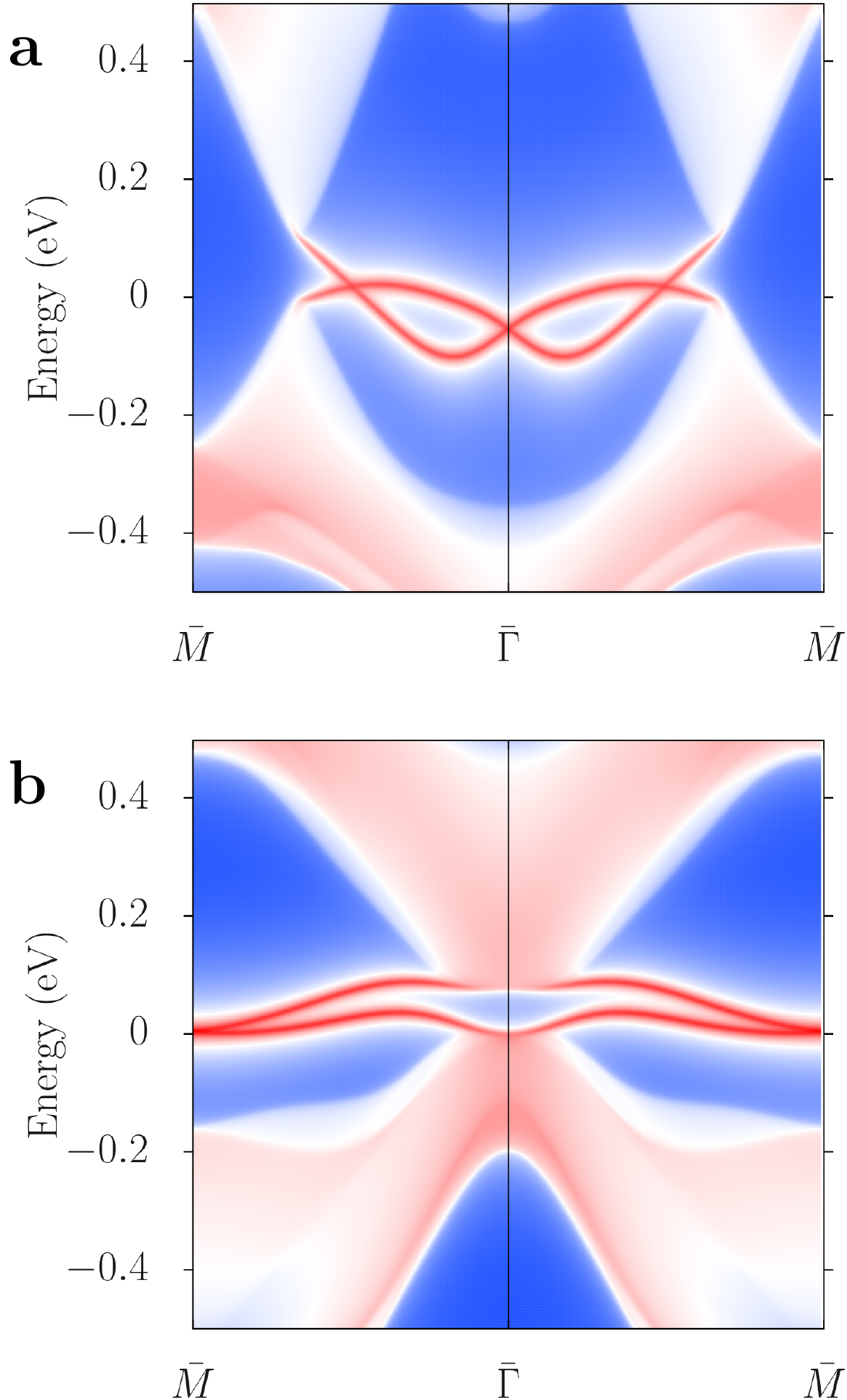}
\caption{\label{fig:edge_bndstr} Topological edge states in the intercalated epitaxial graphene. 
The edge state dispersion of (a) Re-intercalated and 
(b) Ta-intercalated epitaxial graphene. 
}
\end{figure}

The existence of the topological boundary states is an important feature of
a topological system.
In general, topological boundary states are supposed to appear at a boundary
between a topologically nontrivial system and a trivial one
as far as the related symmetry (in our case the time-reversal symmetry) is not broken.
The topological edge states are guaranteed to appear in the bulk energy gap
in such a way that they cross the Fermi level at an odd number of points in half the BZ,
connecting the bulk conduction and valence bands.
They are robust conducting channel since the back-scattering is forbidden
as far as the time-reversal symmetry is kept~\cite{qi2011}.
To calculate the dispersion of the boundary states, we construct Wannier functions
from our DFT calculations and obtain the corresponding Hamiltonian.
Then the edge state dispersion is calculated using the iterative 
Green's function method~\cite{guinea1983,sancho1984,sancho1985,wu2018}
which gives energy- and momentum-resolved local density of states at the edge layers
in the semi-infinite edge structure.
The edge band structures are presented in Fig.~\ref{fig:edge_bndstr}.
In both Re and Ta intercalation cases, we find that the edge states cross the Fermi energy 
(or any horizontal line in the bulk gap) at an odd number of crossing points 
along $\bar{\Gamma}$---$\bar{M}$,
as dictated by the bulk-boundary correspondence.
This confirms that the bulk band topology is nontrivial in accordance with
our $\mathbb{Z}_2$ invariant calculations.

In conclusion, we showed that 5$d$ transition metal intercalation induces
the topological band gap in the epitaxial graphene on the SiC substrate.
It was found that the Re-intercalated graphene has linear Dirac cones with the sizable
SOC-induced gap and the topological character of the occupied bands was
verified by the nontrivial $\mathbb{Z}_2$ invariant.
In the Ta-intercalated graphene, we found the quadratic bands with
the topological band gap.
The broken inversion symmetry and the SOC induced the notable spin-splitting in both systems.
Experimentally, the electronic band structures of the intercalated epitaxial graphene could be 
measured by ARPES~\cite{gierz2010,watch2013,watch2012,emtsev2011,starke2012}, 
and the existence of the topological edge states could be validated 
via transport measurements~\cite{konig2007}.
Our results open a new avenue for the topological phase control of graphene 
and are potentially useful for spintronic applications of graphene and quantum computations.

~\\
\begin{acknowledgments}
This work was supported by the U.S. Department of Energy (DOE), Office of Science, Basic Energy Sciences, Materials Sciences and Engineering Division. The research was performed at Ames Laboratory, which is operated for the U.S. DOE by Iowa State University under Contract No. DE-AC02-07CH11358. Computations were performed through the support of the National Energy Research Scientific Computing Center, which is a DOE Office of Science User Facility operated under Contract No. DE-AC02-05CH11231.
\end{acknowledgments}

\end{document}